\documentclass[twocolumn,
reprint,
showpacs,
preprintnumbers,
aps,
pra,
letterpaper,
superscriptaddress,
nofootinbib,
tightenlines,
floats,
floatfix]{revtex4}
\usepackage{graphicx}
\usepackage{dcolumn}
\usepackage{bm}
\usepackage{latexsym}
\usepackage{amsmath,amssymb}
\usepackage[draft=false]{hyperref}
\usepackage[latin1]{inputenc}
\usepackage{latexsym}
\usepackage{amsmath}
\usepackage{ifpdf}

\begin{document}

\title{Dual atomic interferometer with a tunable point of minimum magnetic sensitivity}

\author{S. Hamzeloui, D. Mart\'{i}nez, V. Abediyeh, N. Arias, E. Gomez}

\email{egomez@ifisica.uaslp.mx}\affiliation{Instituto de F\'{i}sica, Universidad Aut\'{o}noma de San Luis Potos\'{i}, San Luis Potos\'{i} 78290, M\'{e}xico}

\author{V.M. Valenzuela}
\address{Facultad de Ciencias F\'isico Matem\'aticas, Universidad Aut\'{o}noma de Sinaloa, M\'{e}xico}

\date{\today}

\begin{abstract}
Atomic interferometers are often affected by magnetic field fluctuations. Using the clock transition at zero magnetic field minimizes the effect of these fluctuations. There is another transition in rubidium that minimizes the magnetic sensitivity at 3.2 Gauss. We combine the previous two transitions to obtain minimum magnetic sensitivity at a tunable magnetic field between 2.2 and 3.2 Gauss. The two interferometers evolve independently from each other and we control the magnetic sensitivity by changing the population in both transitions with a microwave pulse.
\end{abstract}


\maketitle


\section{Introduction}

The response of atoms to magnetic fields has been an active topic of research for many years. Atoms can be configured as magnetometers that achieve sensitivity slightly better than other available technologies \cite{sheng13}. These sensors have been used to achieve magnetic noise free regions \cite{afach14,baker06,koch15,belfi10}, to study biomagnetism \cite{xia06,hamalainen93}, to generate squeezing in spin systems \cite{schleier10,kuzmich00,sewell12} or to look for physics beyond the Standard Model \cite{smiciklas11,baron14}.

The magnetic sensitivity becomes a source of noise in experiments that use atoms as sensors for other quantities. Magnetic fluctuations are a common source of decoherence in quantum information applications \cite{hammerer10,treutlein06}. The choice of the proper transition and the use of active feedback or magnetic shielding helps minimizing these noise contributions \cite{afach14,ringot01,dedman07}. Atomic clocks, for example, use the so called clock transition that connects levels with no linear Zeeman effect. For atoms trapped with optical beams, there is in addition a differential ac Stark shift between the hyperfine levels that can be canceled by having elliptical polarization at a particular magnetic field value \cite{lundblad10,dudin10,derevianko10,chicireanu11}. Coherence times of six hours has been obtained using rare earth doped crystals at a Zero First Order Zeeman (ZEFOZ) point \cite{zhong15}. The effect of environmental fluctuations is further reduced by the use Dynamical Decoupling (DD) \cite{zhong15,dudin10,heinze13}. The residual magnetic field noise can be characterized using a co-magnetometer while the measurement takes place \cite{harris99,baker06,smiciklas11}. 

To have minimum magnetic sensitivity, the two levels involved in the transition must have the same magnetic response. This is the case for the hyperfine clock transition between $\left| F_1,m_1=0 \right \rangle$ and $\left| F_2,m_2=0 \right \rangle$ at zero field. Bosonic alkali atoms have another magnetically insensitive transition at low magnetic field between $\left| F_1,m_1=-1 \right \rangle$ and $\left| F_2,m_2=1 \right \rangle$ \cite{harber02}. The linear Zeeman shift of the transition cancels, leaving only a quadratic dependence around a particular magnetic field value. The minimum sensitivity for this two photon transitions is achieved at 3.2 Gauss in the case of rubidium \cite{harber02}. Atoms in those two levels have a similar magnetic moment at that field, and it is possible to overlap them in a magnetic trap \cite{matthews98}. Coherence times up to 58 s have been achieved in this transition thanks to the spin self-rephasing mechanism \cite{deutsch10}. Fermionic alkali atoms do not have a clock transition, but they still have a magnetically insensitive single photon transition at low magnetic field between $\left| F_1,m_1=1/2 \right \rangle$ and $\left| F_2,m_2=-1/2 \right \rangle$ \cite{sheng10}.

It is possible to modify the magnetic response of the atoms using microwave dressing. An off-resonant microwave field shifts the point of minimum magnetic sensitivity and, if properly tuned, it is even possible to cancel the quadratic dependence on the magnetic field \cite{sarkany14}. This microwave dressing has been exploited to control the spinor dynamics in Bose Einstein condensates \cite{gerbier06}, increase the coherence times of Qubits \cite{timoney11} and improve the performance of atomic clocks \cite{zanon12}.

In the present work we combine the clock and two photon magnetically insensitive transitions in rubidium to have minimum magnetic sensitivity at a tunable point between 2.2 and 3.2 Gauss. The point of minimum sensitivity depends on the fraction of atoms in each transition. Both interferometers evolve almost simultaneously and they contribute independently to the total signal. Simultaneous interferometers have been used in gravimetry applications to eliminate common noise coming from vibrations \cite{mcguirk02,bonnin13,meunier14} or to determine magnetic field gradients \cite{hu11,wood15}. This has been essential for precision measurements such as the determination of the Newtonian gravitational constant \cite{rosi14}. These works obtain the phase of each interferometer independently in order to eliminate common noise. In our case we excite two transitions on the same atom and we measure directly the combined signal from both interferometers that has the magnetic noise cancellation built in.

\section{Interferometric signal from a dual interferometer}

\begin{figure}[htb!]
\centering
\includegraphics[width=8cm]{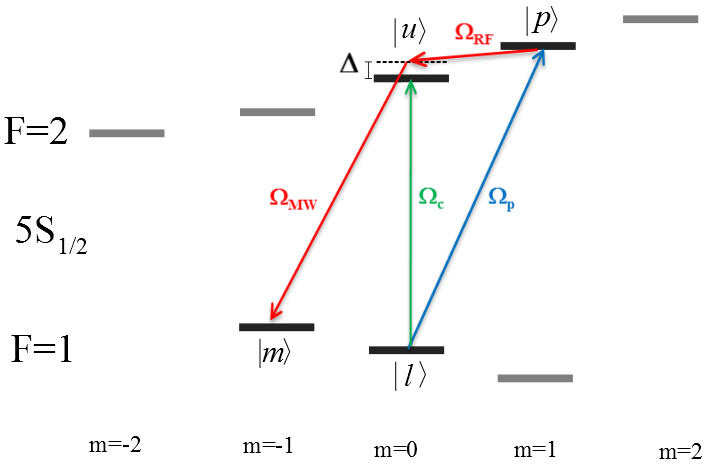}
\caption{\label{rblevels} Energy levels of the ground state 5S$_{1/2}$ of $^{87}$Rb. The green arrow is the clock transition ($\Omega_c$) and the red are the two photon transition ($\Omega_{RF}$ and $\Omega_{MW}$). They are connected by the preparation pulse in blue ($\Omega_p$).}
\end{figure}

We excite two transitions sequentially to obtain two independent interferometers in $^{87}$Rb. The first interferometer works on the clock transition that goes between the hyperfine levels $\left| F=1,m=0 \right \rangle \rightarrow \left| l \right \rangle$ and $\left| F=2,m=0 \right \rangle \rightarrow \left| u \right \rangle$ (green arrow in Fig. \ref{rblevels}), with an associated Rabi frequency $\Omega_c$. The second interferometer works between the hyperfine levels $\left| F=1,m=-1 \right \rangle \rightarrow \left| m \right \rangle$ and $\left| F=2,m=1 \right \rangle \rightarrow \left| p \right \rangle$ (red arrows in Fig. \ref{rblevels}), that we call from now on the two photon transition. We excite this transition by combining a microwave field ($\Omega_{MW}$) and an RF field ($\Omega_{RF}$) detuned by $\Delta$ from the intermediate level to give a two photon Rabi frequency $\Omega_2 = \Omega_{MW} \Omega_{RF} / 2\Delta$ \cite{moler92}.

We initialize the atoms in $\left| l \right \rangle$ and we transfer some of the population to $\left| p \right \rangle$ with the preparation pulse ($\Omega_p$) shown in blue in Fig. \ref{rblevels}. The pulse duration ($\tau$) determines the fractional population of atoms in the two photon ($P= \sin^2{[\Omega_p \tau / 2]}$) and clock ($1-P$) interferometers. The state after the preparation pulse is given by

\begin{equation}
\left| \Psi \right\rangle = \sqrt{1-P} \left| l \right\rangle -i \sqrt{P} \left| p \right\rangle.
\label{startpsi}
\end{equation}
We apply a short resonant $\pi$/2 pulse on the two photon transition followed by a similar pulse on the clock transition to initialize the two interferometers. After some evolution time ($T$) we obtain

\begin{eqnarray}
\left| \Psi \right\rangle & = & \sqrt{(1-P)/2} e^{i \phi_1} \left( \left| l \right\rangle -i e^{i \delta_{1} T} \left| u \right\rangle \right) \nonumber \\
& & -i \sqrt{P/2} e^{i \phi_2} \left( \left| p \right\rangle - i e^{i (\delta_2 [T + T_p] + \phi_3)} \left| m \right\rangle \right),
\label{psi2}
\end{eqnarray}
with $\delta_j = \omega_{mj} - \omega_{aj}$ the detuning, $\omega_{mj}$ and $\omega_{aj}$ the microwave and atomic frequencies of the clock ($j=1$) and two photon ($j=2$) transitions. The field that is resonant for one transition has a large detuning for the other transition and introduces a negligible light shift indicated by $\phi_1$, $\phi_2$ and $\phi_3$. The free evolution time of both interferometers differ by the duration of the clock transition pulse ($T_p$) which is much smaller than the time between pulses ($T$). Temporal fluctuations of the microwave phases will introduce noise given the non simultaneous excitation of both interferometers. 

For small magnetic fields the frequency separation between the levels $\left| F=1,m_1 \right \rangle$ and $\left| F=2,m_2 \right \rangle$ is given by \cite{castanos14}

\begin{eqnarray}
\omega_{ai} & = & \omega_{HFS} \left[ 1 + \left( \frac{m_1 + m_2}{4}+\gamma_2 (m_2-m_1) \right) aB \right. \nonumber \\
& & \left. + \left( \frac{1}{2}-\frac{m_1^2+m_2^2}{16} \right) a^{2}B^{2} \right],
\label{maclaurin}
\end{eqnarray} 
where $\omega_{HFS} / 2 \pi = 6.834$ GHz is the hyperfine splitting, $\gamma_2=(g_{I} m_e / m_p)/ (g_s + g_I m_e/m_p)=-4.97 \times 10^{-4}$, $a=(g_s \mu_B + g_I \mu_N)/ \hbar \omega_{HFS}=4.1 \times 10^{-4}~{\rm G}^{-1}$, $g_{s}$ and $g_{I}$ are the g-factors of the electron and nucleus, $\mu_{B}$ and $\mu_{N}$ are the Bohr and nuclear magneton, $m_{e}$ and $m_{p}$ are the mass of the electron and proton, and the values have been evaluated for $^{87}$Rb. Equation \ref{maclaurin} gives a detuning for the two transitions of interest

\begin{eqnarray}
& & \delta_1 = (\omega_{m1} - \omega_{HFS} - D_1 B^2) \nonumber \\
& & \delta_2 = (\omega_{m2} - \omega_{HFS} - C_2 B - D_2 B^2), 
\label{phaseaq}
\end{eqnarray}
with

\begin{eqnarray}
& & D_1 = \frac{a^2}{2} \omega_{HFS} = 2 \pi ~ (575 ~ {\rm Hz/G}^2) \nonumber \\
& & C_2 = (2 a \gamma_2) \omega_{HFS} = 2 \pi ~ (-2785 ~ {\rm Hz/G}) \nonumber \\
& & D_2 = \frac{3 a^2}{8} \omega_{HFS} = 2 \pi ~ (431 ~ {\rm Hz/G}^2).
\label{coefsdelta}
\end{eqnarray}

A second resonant $\pi$/2 pulse completes the interferometer and we measure the fraction of atoms in the upper hyperfine level. The detection method does not distinguish between Zeeman sublevels in the upper hyperfine level giving a signal equal to

\begin{eqnarray}
S = \left| c_u \right| ^2 + \left| c_p \right| ^2 & = & \frac{1}{2} \left[ 1 + (1-P) \sin{(\delta_1' T)} \right. \nonumber \\
& & \left. + P \sin{(\delta_2' T + \Phi)} \right].
\label{signal2}
\end{eqnarray}
This expression shows the sum of the fringes from both transitions. In this expression we added an offset on the detuning $\delta_1' T= \delta_1 T + \pi / 2$ and $\delta_2' T= \delta_2 T - \pi / 2$ to have the fringes of both interferometers in phase. The fringes on the second interferometer are shifted by $\Phi=2 \delta_2 T_p + 2 \phi_3$. At the small detunings of interest ($\delta_2 \sim \pi/2 T$) the first term gives a negligible contribution ($2 \delta_2 T_p \sim \pi T_p/T$) since in our case $T_P/T \ll 1$. The differential light shift contribution is also negligible ($\phi_3 < \pi (6 \times 10^{-4})$), therefore we neglect the shift $\Phi$ from now on. The two interferometers work independently and do not interfere with each other. The preparation pulse connects both transitions at the beginning, but there is no similar pulse at the end to make them interfere. This is why the relative phase between the two interferometers ($\phi_1$ and $\phi_2$) in Eq. \ref{psi2} does not appear in Eq. \ref{signal2} . The above calculation assumes resonant fields during the excitation and considers the effect of the detuning only during the free evolution time ($T$). This is a good approximation when the time between pulses is much larger than their duration.

To have magnetic field insensitivity we must have $\partial S / \partial B = 0$. Care must be taken to avoid removing at the same time the sensitivity to the quantity of interest. For instance, suppose you are interested in measuring time. If you combine two interferometers with opposite phase, then you eliminate the fringes and therefore you remove the magnetic sensitivity, but at the same time you are also no longer able to measure time.  Consider the case where we populate the two interferometers equally ($P=1/2$). Each interferometer separately would give fringes with half the total visibility. The two interferometers must be set in phase to have complete sensitivity to the quantity of interest (time in the example above), but with an opposite response to magnetic field fluctuations. In this case a variation in magnetic field would shift the clock fringes slightly in one direction and the two photon fringes in the opposite direction. An interferometer operating at the middle of the fringe would remain at the same position with the magnetic variation, minimizing the sensitivity to fluctuations.

The signal variation $\Delta S = (\partial S / \partial B) \Delta B$ for a small change in the magnetic field $\Delta B$ around $B_0$ can be calculated from Eq. \ref{signal2} for an interferometer operating at the middle of the fringe ($\delta_1',\delta_2' \approx 0$), and gives

\begin{equation}
\Delta S = \left( -2 D_1 B_0 + P \left[ 2 (D_1 - D_2) B_0 - C_2 \right] \right) T \Delta B.
\label{eqdeltas}
\end{equation}
The minimum magnetic sensitivity ($\Delta S = 0$) is achieved at

\begin{equation}
B_{min}= \frac{-P C_2}{2 P D_2 + 2 (1-P) D_1}.
\label{minB}
\end{equation}
We see from this equation that the magnetic field value of minimum sensitivity can be in principle continuously tuned between 0 and 3.2 Gauss by varying the fraction of atoms in each transition. In particular if all the atoms are in the clock interferometer, then $P=0$ and the minimum sensitivity happens at 0 Gauss as expected. If instead all the atoms are in the two photon interferometer, then $P=1$ and we get the well known $B_{min}=-C_2 / 2 D_2=3.2$ Gauss \cite{harber02}. In the next sections we provide the experimental demonstration of this dual interferometer with tunable point of minimum magnetic sensitivity.

\section{Experimental setup}

We start with 10$^{8}$ atoms captured in a Magneto Optical Trap (MOT) \cite{valenzuela12}. We apply an optical molasses to the atoms during 3 ms to lower their temperature to 3 $\mu$K \cite{hamzeloui16}. To initialize all the atoms in $\left| l \right \rangle$ (Fig. \ref{rblevels}), we simultaneously excite them with a depumper beam on the 5S$_{1/2}$ $F=2$ to 5P$_{3/2}$ $F=2$ transition and a $\pi$ polarized beam on the 5S$_{1/2}$ $F=1$ to 5P$_{3/2}$ $F=1$ during 100 $\mu$s. The optical pumping process puts more than 94\% of the atoms in the desired state. We block the beams with mechanical shutters to ensure that there is no scattered light during the interferometric sequence \cite{martinez11}.

There is a magnetic field of 380 mG during optical pumping with the magnetic gradient off. We quickly change the magnetic field to a particular desired value using two switches (Apendix) while the atoms are in free fall. We allow 7 ms for the magnetic field to stabilize before we apply the interferometer sequences. The magnetic field changes less than 25 mG during the time between the interferometry pulses with a reproducibility better than 1 mG (Fig. \ref{igbtimage}).

\begin{figure}[htb!]
\centering
\includegraphics[width=8cm]{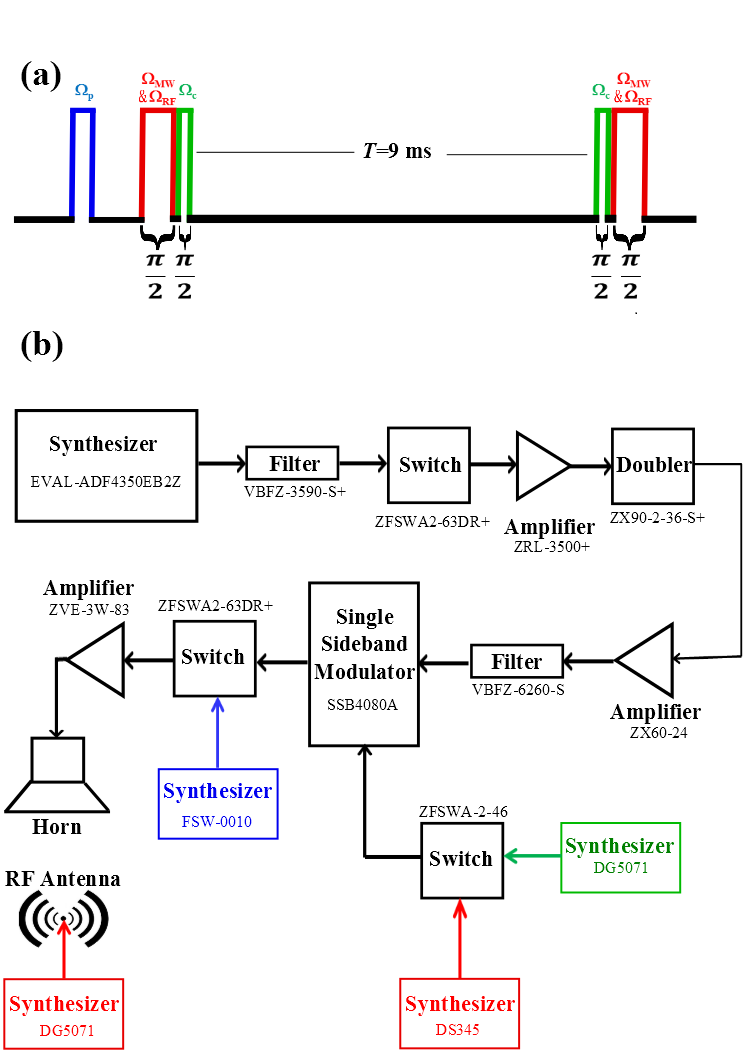}
\caption{\label{micsequence} (a) Microwave pulse sequence and (b) microwave system (b). The colors correspond to those of the transitions in Fig. \ref{rblevels}.}
\end{figure}

Fig. \ref{micsequence} depicts the excitation sequence (a) and the microwave system (b). The preparation pulse determines the population on each interferometer (blue in Figs. \ref{rblevels} and \ref{micsequence}). It occurs after the initialization steps at 380 mG, right before ramping up the magnetic field and it has a Rabi frequency of $\Omega_p / 2 \pi = 2.5$ kHz. It is generated by a synthesizer (Phase Matrix FSW-0010) with an internal switch to control the pulse timing and duration. An external switch selects between this and the other microwave signals. The microwave pulses for the clock interferometer are generated by combining a fixed high frequency signal with a tunable low frequency one in a single sideband modulator (green in Figs. \ref{rblevels} and \ref{micsequence}). The high frequency signal comes from a PLL synthesizer (EVAL-ADF4350EB2Z) that is frequency doubled, filtered and amplified \cite{valenzuela13}. The microwave field of the two photon interferometer ($\Omega_{MW}$) is generated in a way similar to the clock transition but with an independent low frequency generator (red in Figs. \ref{rblevels} and \ref{micsequence}) that is selected through a switch. The RF part of the two photon transition ($\Omega_{RF}$) comes directly from an RF generator that feeds a resonant loop homemade antenna that is located near the position of the atoms.

To ensure frequency stability, all the synthesizers are locked to an atomic clock (SRS FS-725). All the microwave signals go through a common amplifier and they are emitted from a horn. Undesired frequencies in the single sideband modulator are suppressed by more than 36 dB and have a negligible effect on the atoms since they lie 14 MHz away from the clock transition.

We collect the fluorescence from the atoms in a double relay imaging system with an iris in the middle for background reduction \cite{hamzeloui16}. The light goes to a CCD camera and to a photo diode connected to a data acquisition card. After the interferometer sequence we determine the fraction of atoms in the $F=2$ hyperfine level by shining a beam resonant with the cycling transition and normalizing it with the signal obtained after repumping all the atoms to the $F=2$ level.

\section{Rabi oscillations in the two transitions of interest}

To characterize the clock transition we do not apply any preparation pulse ($\Omega_p$) so that all the atoms remain in $\left| l \right\rangle$ (Fig. \ref{rblevels}). The horn orientation must be selected to be able to drive both the $\pi$ and $\sigma^+$ transitions needed for the clock and two photon transitions. The Rabi frequency for the clock transition is $\Omega_c / 2 \pi =4.5$ kHz giving $\pi/2$ pulses of 55 $\mu$s.

We induce transitions between Zeeman sublevels to characterize the RF part of the two photon transition ($\Omega_{RF}$). Starting in $\left| l \right\rangle$, we apply a two step transition, first to $\left| F=1,m=1 \right\rangle$ with the RF and then to $\left| F=2,m=1 \right\rangle$ with a microwave pulse, to detect them on the cycling transition. We adjust the magnitude of the RF field to have a Rabi frequency of $\Omega_{RF} / 2 \pi =6$ kHz, similar to that obtained for $\Omega_{MW}$.

\begin{figure}[htb!]
\centering
\includegraphics[width=8cm]{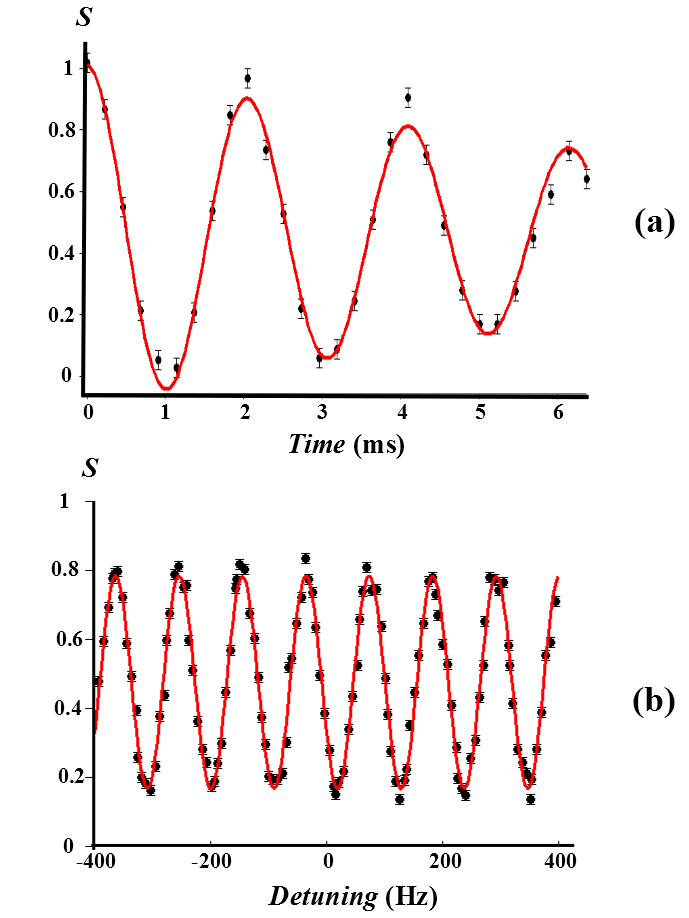}
\caption{\label{twophotonrabi} (a) Rabi oscillations and (b) interference fringes for the two photon transition.}
\end{figure}

We apply a $\pi$ preparation pulse ($\Omega_p$) to send all the atoms to $\left| p \right\rangle$ to characterize the two photon transition. The detuning ($\Delta / 2 \pi = 70$ kHz) is sufficiently large to avoid excitations to the intermediate level. Figure \ref{twophotonrabi}a shows the Rabi oscillations for the two photon transition, with a typical Rabi frequency of $\Omega_{2} / 2 \pi =0.5$ kHz giving $\pi/2$ pulses of 500 $\mu$s. The decay of the oscillations is most likely due to slightly inhomogeneous microwave illumination of the atoms. The interference fringes for the two photon transition are shown in Fig. \ref{twophotonrabi}b for a time between pulses of $T=9$ ms and they have a small shift with respect to Eq. \ref{signal2}.

\begin{figure}[htb!]
\centering
\includegraphics[width=8cm]{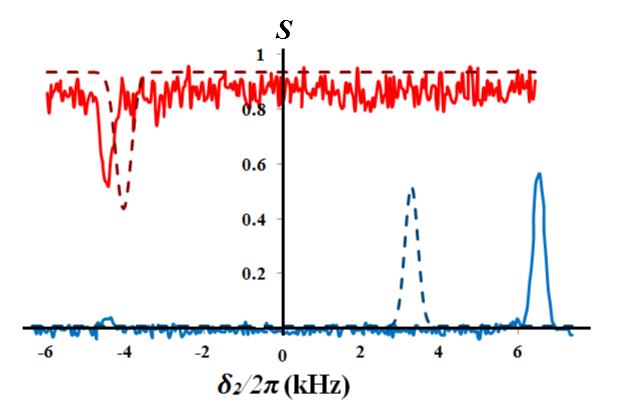}
\caption{\label{crossspectrum} Spectrum taken at 3.2 Gauss with only the two photon fields on ($\Omega_{MW}$ and $\Omega_{RF}$) with all the atoms initially in $\left| l \right\rangle$ (lower solid blue) and $\left| p \right\rangle$ (upper solid red) levels. The dashed lines correspond to fits to data taken at 2.24 Gauss under the same conditions. The red peaks correspond to the desired two photon transition and the blue peaks are the undesired two photon excitation of the clock transition.}
\end{figure}

To treat each interferometer independently, it is important to minimize the effect of the clock microwave field on the two photon transition and also the other way around. Having the fields for both transitions simultaneously on could potentially drive undesired multiphoton transitions. Instead we shine the two fields sequentially (Fig. \ref{micsequence}). We leave the clock interferometer pulses inside the two photon ones since they are shorter (higher $\Omega_c$) and also because the microwave field for the clock transition alone cannot drive the two photon transition. In contrast, it is possible to drive the clock transition with a two photon transition. Figure \ref{crossspectrum} shows a two photon spectrum taken at 3.2 G with the preparation pulse off so that all the atoms are in level $\left| l \right\rangle$ (lower solid blue curve), and with a $\pi$ preparation pulse with all the atoms in $\left| p \right\rangle$ (upper solid red curve). At this magnetic field the desired two photon transition ($\left| p \right\rangle \rightarrow \left| m \right\rangle$, red left peak) is spectrally resolved from the undesired two photon excitation of the clock transition ($\left| l \right\rangle \rightarrow \left| u \right\rangle$, blue right peak), thanks to the quadratic Zeeman shift. The separation between peaks is 21 times larger than their width, with this last one determined by the pulse duration. The two photon excitation of the clock transition gives a negligible population transfer of at most 0.3 \%. The height of the peaks depends on the pulse duration that is shorter than a $\pi$ pulse in this case. There is an small peak in the lower blue trace of Fig. \ref{crossspectrum} at -4.5 kHz coming from the residual population in the $\left| F=1,m=-1 \right\rangle$ level after optical pumping. The dashed lines in Fig. \ref{crossspectrum} correspond to the fit of the data taken at 2.24 G, and show that the peaks get closer as we reduce the magnetic field.

\section{Dual interferometer}

\begin{figure}[htb!]
\centering
\includegraphics[width=8cm]{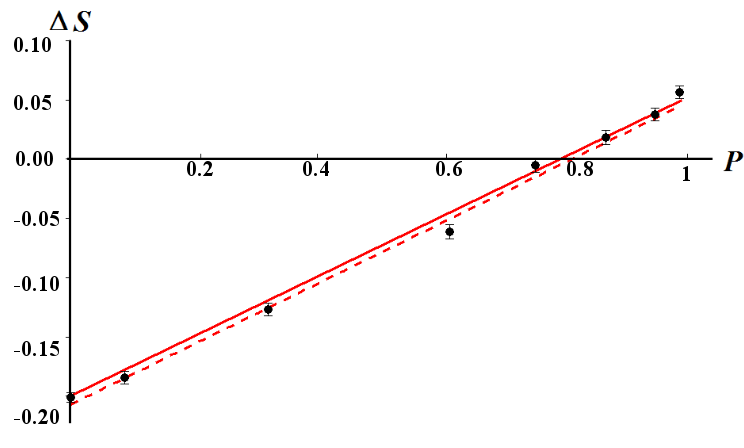}
\caption{\label{signalpopulation} Signal variation with magnetic field ($\Delta S$) as a function of the fraction of atoms in the two photon transition transition ($P$) at 2.5 G. The solid red line is a fit to the data and the red dashed line gives the theoretical signal (Eq. \ref{eqdeltas}) for $\Delta B$=1.2 mG.}
\end{figure}

We excite both interferometers (clock and two photon) using the sequence from Fig. \ref{micsequence} to demonstrate the tunable point of minimum magnetic insensitivity. We scan the frequency of each interferometer independently to obtain fringes and we sit at the middle of the fringe in the rising slope in both of them. To characterize the magnetic sensitivity at a particular magnetic field $B_0$, we measure the interferometer signal (Eq. \ref{signal2}) with a magnetic field slightly higher and lower and we take the difference of the two, $\Delta S = S (B_0 + \Delta B /2) - S (B_0 - \Delta B /2)$. Figure \ref{signalpopulation} shows a plot of $\Delta S$ as we vary the fractional population in the two photon ($P$) and clock ($1-P$) interferometers by changing the preparation pulse duration ($\tau$) at 2.5 G. When all the atoms are in the clock transition ($P=0$) the fringes shift in one direction and $\Delta S<0$, and the opposite happens with all the atoms in the two photon transition ($P=1$). There is a population in between ($P=0.80 \pm 0.03$) where there is no shift, and at this point we get minimum magnetic sensitivity.

The theoretical signal variation (Eq. \ref{eqdeltas}) is also shown as a red dashed line in Fig. \ref{signalpopulation} for $\Delta B$=1.2 mG. This $\Delta B$ value gives good agreement with the data and it is similar to that estimated from spectroscopic measurements. It would introduce a shift of 3.1 \% (-0.7 \%) of a fringe for the clock (two photon) transition. The visibility remains almost unchanged with this small shift and the interferometer retains full sensitivity to other quantities of interest. The sensitivity suppression in Fig. \ref{signalpopulation} depends on how close we can be to $P=0.8$. In our case we control the population to $\Delta P \sim \pm 0.05$, giving us a factor of 4 (20) sensitivity improvement at the 2.5 G of Fig. \ref{signalpopulation} with respect to only using the two photon (clock) transition. In our measurements we work at a particular $B_0$ and we vary $P$ since that is more stable. Keeping $P$ fixed at 1 and instead changing $B_0$ gives a minimum magnetic sensitivity at $3.20 \pm 0.02$ G as expected, but involves a more complicated experimental procedure.

\begin{figure}[htb!]
\centering
\includegraphics[width=8cm]{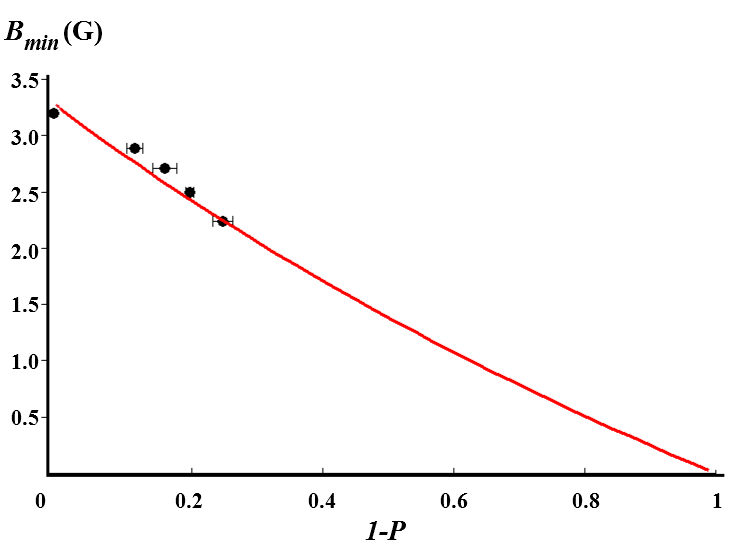}
\caption{\label{bminplot} Fraction of atoms in each interferometer ($P$) required to have minimum sensitivity at a particular magnetic field ($B_{min}$). The solid line corresponds to Eq. \ref{minB}.}
\end{figure}

Figure \ref{bminplot} shows the population fraction ($P$) needed to have minimum magnetic sensitivity at a particular field ($B_0=B_{min}$). The data follows the theoretical prediction (solid line) from Eq. \ref{minB}. The theory works well as long as the two interferometers can be considered independent of each other. As the magnetic field is decreased, the separation between the two peaks in Fig. \ref{crossspectrum} decreases and the two transitions start affecting each other. In particular, one starts having a non negligible population transfer on the clock transition by the two photon field, which becomes a concern for a precision measurement. Keeping this population transfer below 1$\%$ limits the operation of the dual interferometers to lower magnetic fields in Fig. \ref{bminplot}.

The dual interferometer maintains a similar quadratic dependence on magnetic field as the independent interferometers, but shifts the point of minimum sensitivity to the desired value. The shift is almost linear with the population in each interferometer (Fig. \ref{bminplot}) as long as the visibility in both interferometers is similar. The technique presented here has the advantage that the atoms evolve in complete darkness, compared with RF dressing techniques \cite{sarkany14} that requires maintaining a stable RF field on during the free evolution. We don't need to extract the phase of each interferometer separately as in other
common noise suppression techniques with dual interferometers \cite{bonnin13}, instead we just read their
combined signal.

\section{Conclusion}
We have presented a dual interferometer that achieves a minimum magnetic sensitivity at a tunable value of the magnetic field. Combining the clock and the two photon transition between $\left| F=1,m=-1 \right\rangle$ and $\left| F=2,m=1 \right\rangle$ in $^{87}$Rb we were able to tune the point of minimum magnetic sensitivity between 2.2 and 3.2 G by changing the fraction of atoms in each interferometer. The dual interferometer may be useful in applications where low magnetic sensitivity is required at a particular magnetic field.

\textbf{Acknowledgments}

Project supported by CONACyT (Frontiers of Science, Infrastructure, Catedras and the Research for Education Fund), UASLP and Marcos Moshinsky Foundation. We thank Jonathan Espinosa and Jos\'e Rocha for their assistance with electronics.

\section{Appendix}
\begin{figure}[htb!]
\centering
\includegraphics[width=8cm]{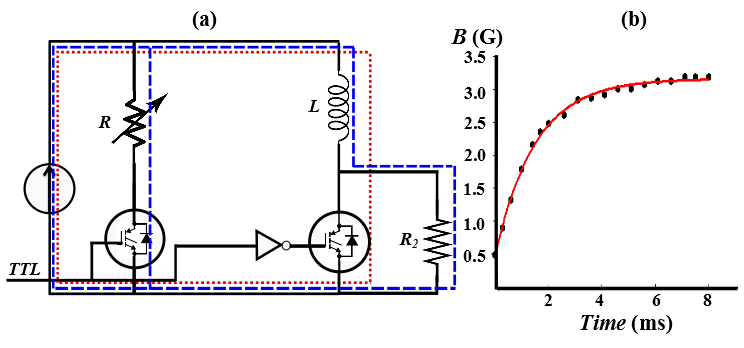}
\caption{\label{igbtimage} (a) Circuit to rapidly switch between the low and high bias magnetic fields. Paths with high (low) TTL are in dashed blue (dotted red). (b) Bias magnetic field as a function of time measured with microwave spectroscopy on the atoms.}
\end{figure}

Here we present the circuit used to rapidly change the magnetic field from 380 mG to any desired value up to 3.2 G. The system uses two Insulated-Gate Bipolar Transistors (IGBT) to redirect part of the current from the coil to a dummy load. The two IGBT's have opposite logic so that one is open when the other one is closed, and they are both controlled with a TTL pulse from the control system. Under the low TTL condition all the current from the supply (that works in constant current mode) goes through the coils producing the desired bias magnetic field (dotted red in Fig. \ref{igbtimage}a). Under the high TTL condition we split the current between the path with the coils and that of a dummy resistor (dashed blue in Fig. \ref{igbtimage}a). The value of the resistor is adjusted to have the 380 mG used during the preparation stage. We measured the magnetic field as a function of time using microwave spectroscopy on the atoms (Fig. \ref{igbtimage}b). The magnetic field approaches exponentially the desired value with a $1.5 \pm 0.1$ ms time constant ($1/e$).

There is a residual magnetic gradient smaller than 30 mG/cm at the position of the atoms from the ion pump. The vertical bias field is produced by two squared shaped Helmholtz coils with a side and separation of 30.5 cm. The coils produce no magnetic gradient at the middle position where the atoms are located. The coil's supply (Agilent E3614A) has a noise of 300 ppm that corresponds to less than 1 mG. All the above produce an average magnetic field felt by the atoms between $\pi/2$ pulses with a reproducibility better than 1 mG.

\end{document}